# A proposal for Transversal Computer-related Strategies & Services for Scientific and Training efforts for the LASF4RI [1]

**Thematic areas**

- Instrumentation and Computing
- Capacity Building

**Arturo Sánchez Pineda ([arturos@cern.ch](mailto:arturos@cern.ch))** Dipartimento di Fisica, Università di Udine, Istituto Nazionale di Fisica Nucleare (INFN) & International Centre for Theoretical Physics (ICTP), Italia. TDAQ-ATLAS at the European Organization for Nuclear Research (CERN).

January 20th, 2020

## Abstract

This schematic proposal is looking to give a first view of the different and vital services, protocols, tools and know-how relative to the Scientific Computing (SC) and Information Technology (IT) for scientific endeavours and capacity-building projects in the Latin America region. The proposal of transversal services and protocols for the design, development, deployment, distribution, training, publication and citation, proper accreditation and dissemination of scientific experiments, data, processes, software, documentation, results and resources using world-leading protocols and industrial standards under the Open Access philosophy is presented. It is showing a dedicated review of scenarios and proposals that can be seen under the umbrella of a "SC+IT Hub". Also, it is presented effective usage of current hybrid spaces (physical and virtual) that contains very well known industrial and academic resources and practical ideas, and how to deploy those for current -diverse- and future experiments and research teams in the region.

---

[1] **The LASF4RI** is calling for inputs on the next steps regarding the coordination of resources for large scientific infrastructures among countries in Latin America ([https://lasf4ri.org](https://lasf4ri.org))

# Introduction

**This White Paper looks to start a discussion and to address as early as possible aspects that are of significance for almost any scientific endeavour in terms of Computation and Science Reproducibility.** The following pages are supported on 10+ years of expertise in Large Experiment Facilities and Open Access projects worldwide and the advice of multiple researchers in the area.

**The core idea** is to let the community know of the most current tools relative to an "Experiment" and "Collaboration" need. Also, the creation of virtual spaces where the researchers and funding agencies can find reliable and updated information/consultancy on SC and related services and tools for their experiments and institutions. In subjects like:

**Experimental assembly** (usually close to the hardware and experimental setups)
- Data acquisition systems
- System on Chip (**SoC**)
- The transmission systems, data pre-processing, triggers
- Data storage and pre-bookkeeping systems
- Selection, classification and distribution of data processes.

**Software-based**
- Data storage system
- Data bookkeeping for human and machine reading
- Data analysis system
- Software development
- Versioning of the software
- Pipelines, Continuous Integration and Continuous Development (CI/CD)
- Use of local, distributed, Cloud, hybrid Academic / Commercial Computing
- Results collection and visualisation systems
- Monitoring and review of mechanism
- Assignment of DOIs and licences
- Training of research team members and other personnel
- Writing reports, documentation, protocols and papers
- Reviews of the documents, proposals and publications
- Input from external advisors and experts.

**As shown below, the ideas are in figures that try to exploit their synthesis power, but with the compromise it will be significantly enhanced and updated to record such resources and practical ideas that researchers can start to implement right away.**
The March 2020 meeting will be the best scenery to show and get proper feedback.

Figure **1** shows a simplified view of multiple experiments and the ways they access to services and tools in the areas of Scientific Computing (SC) and Information Technology (IT).

It illustrates that independent experiments usually get their computer resources, data manager, analysis tools, publication system, etc. in independent and unconnected ways and sources.

This is a usual source of redundancy and multiple-expending for the same type of services and equipment.

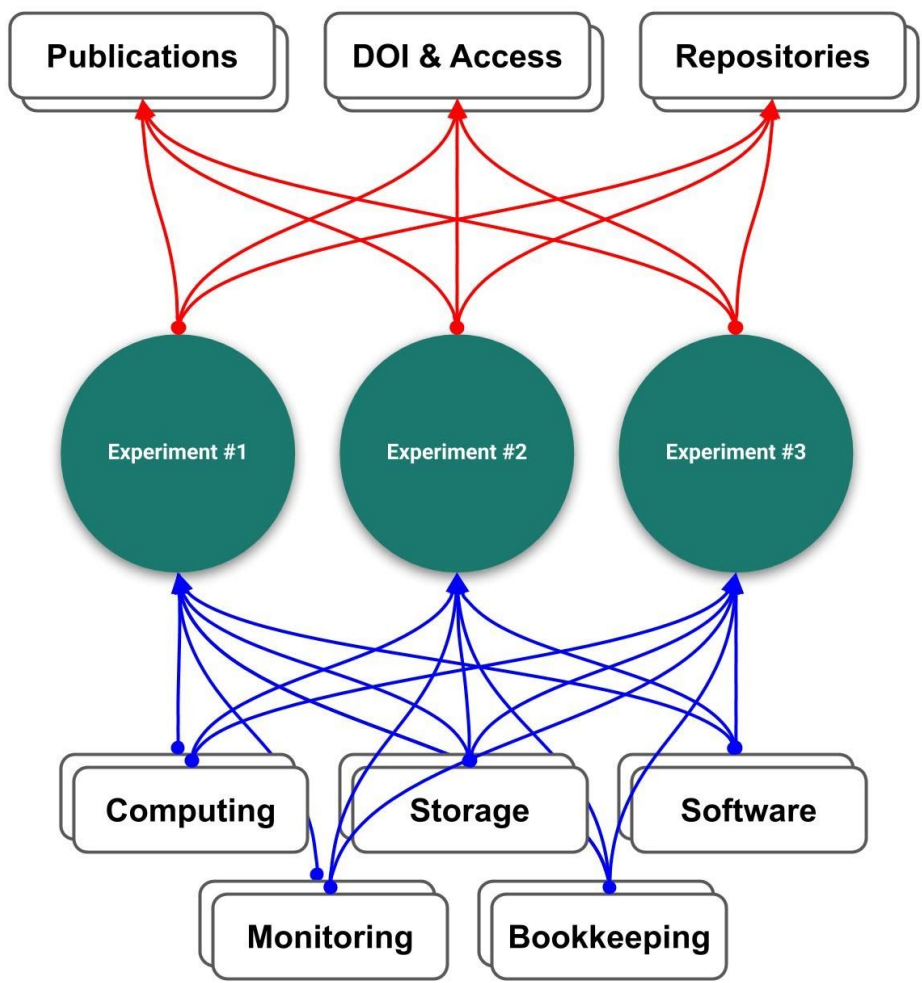

**Figure 1**: A simplified view of multiple experiments and the ways they access to services and tools in the areas of Scientific Computing (SC) and Informatics Technology (IT).

The other set of Figures (2-9) below shows how we can use transversal services to homogenise, integrate, reduce costs and improve the quality of the scientific research and researchers.

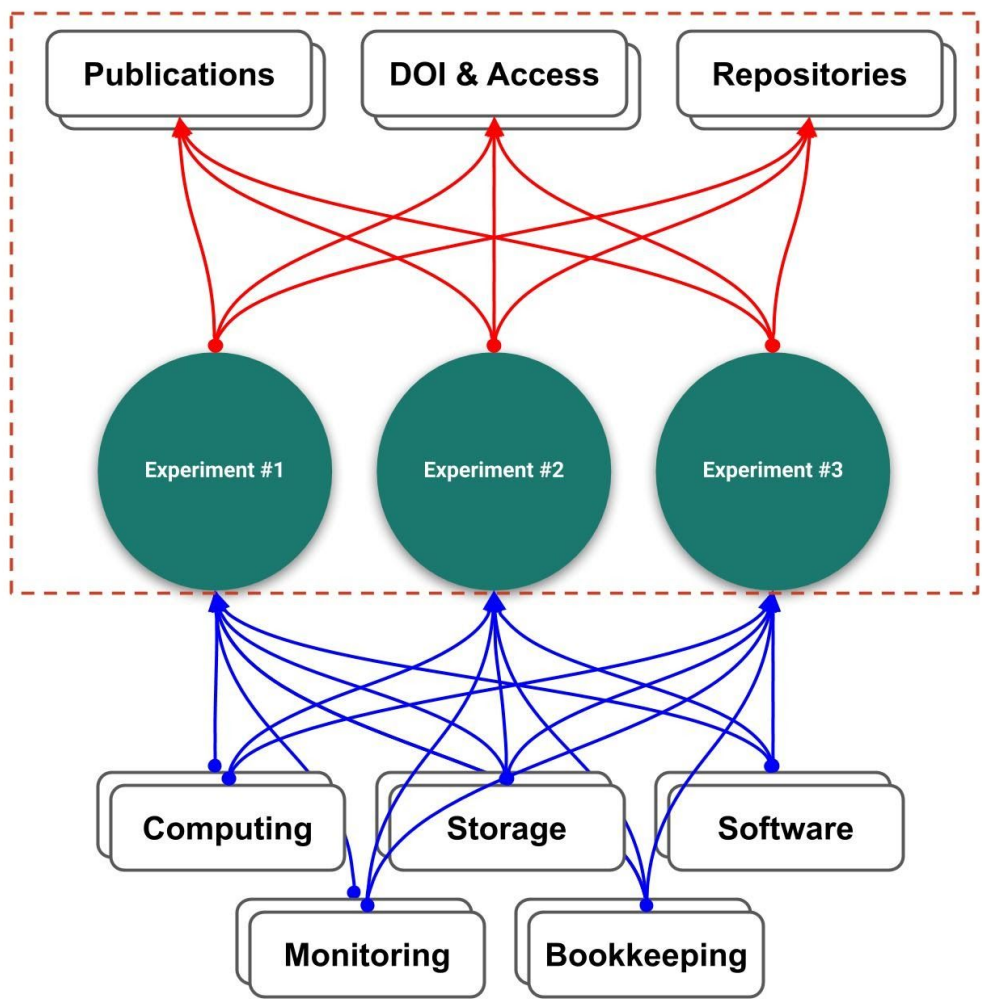

**Figure 2**: A simplified view of multiple experiments and the ways they access to services and tools in the areas of SC and IT highlighting the area of usage of final services.

This approach of maximising the use of common resources when possible -used in multiple and large collaborations- allows the possibility to deploy generic and eventually economic common infrastructure. Single purchases and/or contracts for multiple experiments and groups can help to optimise the use of funding and to get more specialised equipment.

Also, in terms of the visibility to possible external funding agencies, it is much more appreciated, especially in multi-country institutions and universities.

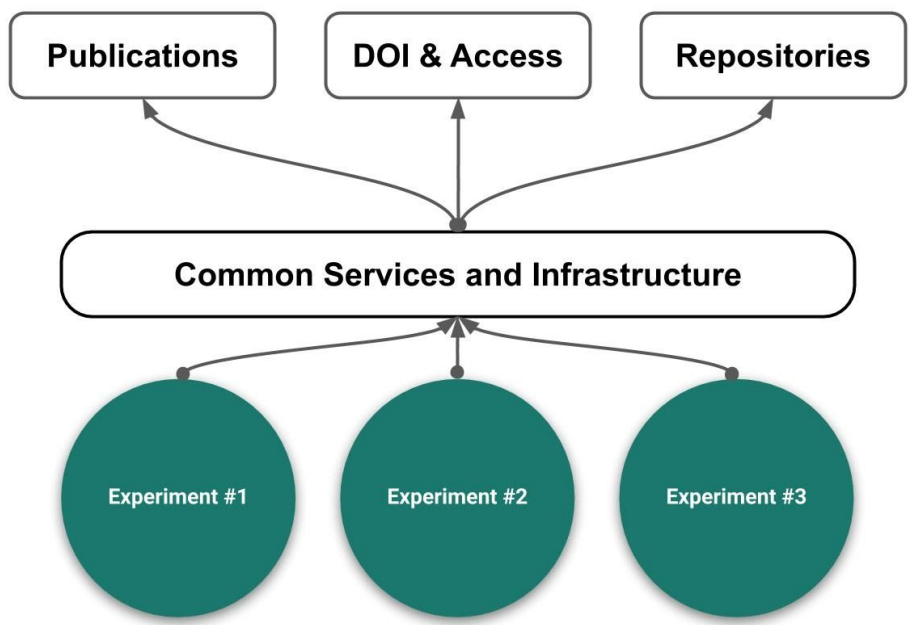

**Figure 3**: A simplified view of multiple experiments and the ways they can access to (final) services and tools in the areas of SC and IT using a common Service Provider Platform.

Many of these common resources are already in place. Open Access, Open Source, Open Data, Open Software and Open Hardware define current and future scientific endeavours. This environment -and legal framework- is beneficial for long terms and fully support Open "X" projects that we can profit and contribute also. So, in a first attend to create a manual of guide for researchers in the region will already have solid technology and resources that are planned to be supported for decades.

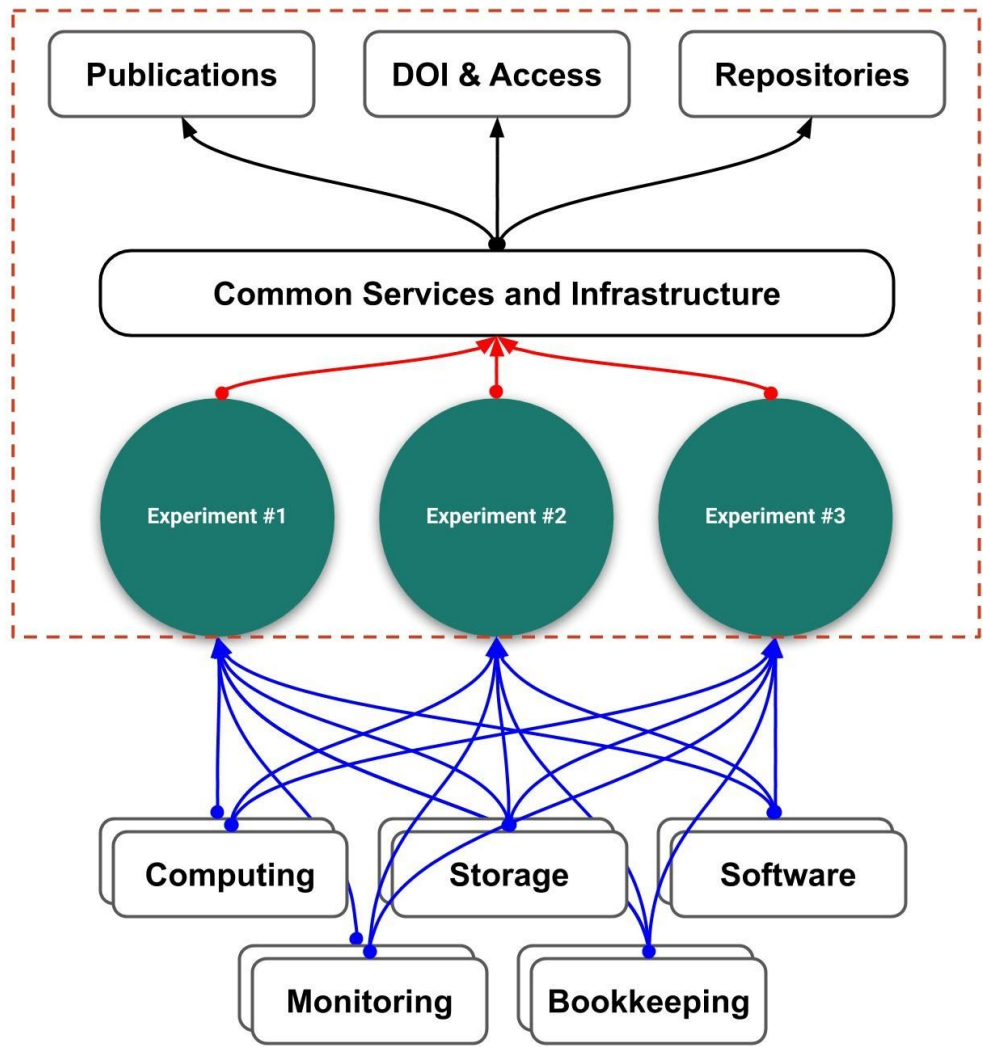

**Figure 4**: A simplified view of multiple experiments and the ways they can access to services and tools through a common Service Provider Platform in the areas of SC and IT highlighting the area of usage of final services.
.

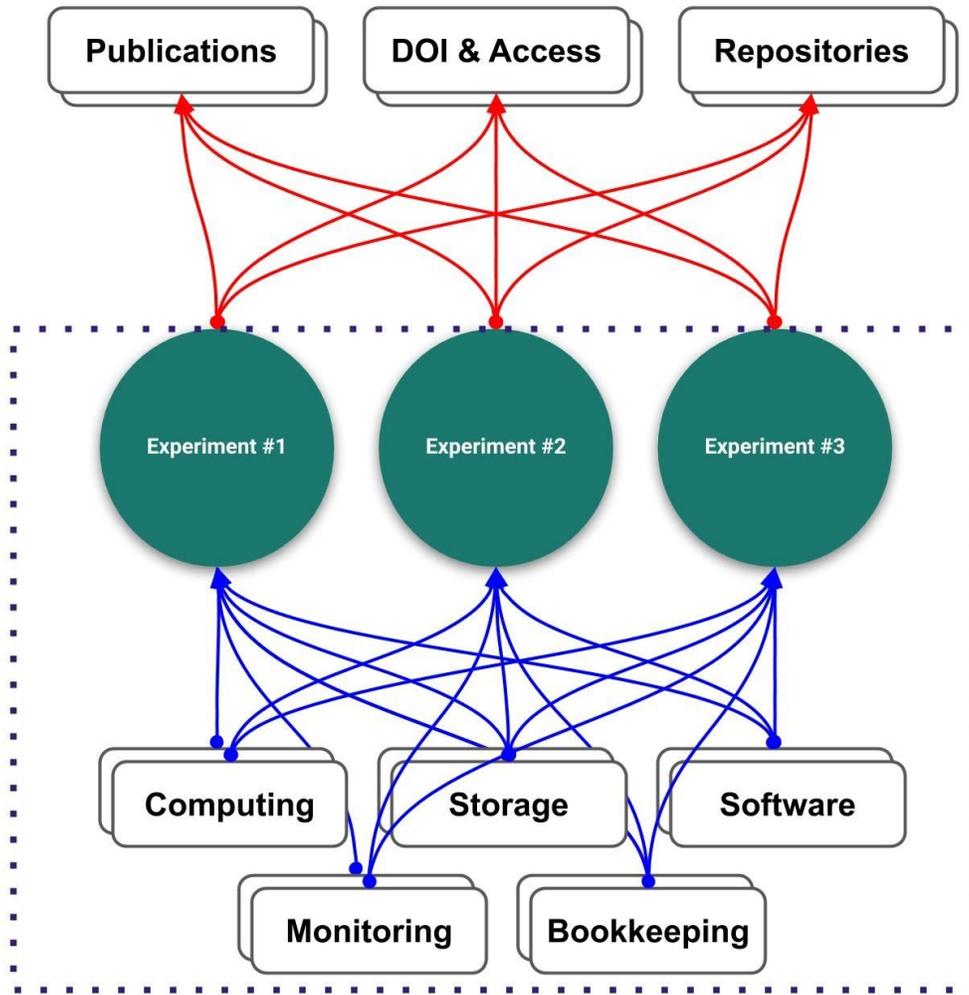

**Figure 5**: A simplified view of multiple experiments and the ways they access to services and tools in the areas of SC and IT highlighting the area of usage of initial-continuous services.

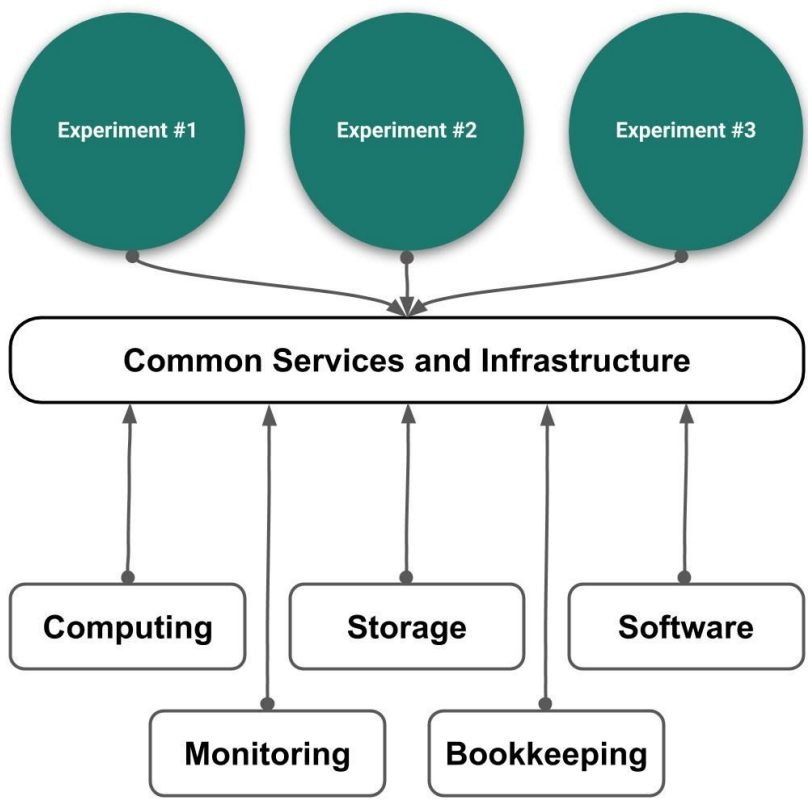

**Figure 6**: A simplified view of multiple experiments and the ways they
can access to (initial-continuous) services and tools in the areas of SC
and IT using a common Service Provider Platform.

The use of transversal common resources and protocols also allow doing more. Many of the mentioned resources are free to be used. And with a community that supports their deployment and can assist in any moment.

Because they are used in the industry, meaning that we are in-line with industrial standards. Keeping a high quality of the activities and allowing the researchers to have the know-how that can be relevant in their professional future inside and outside academia.

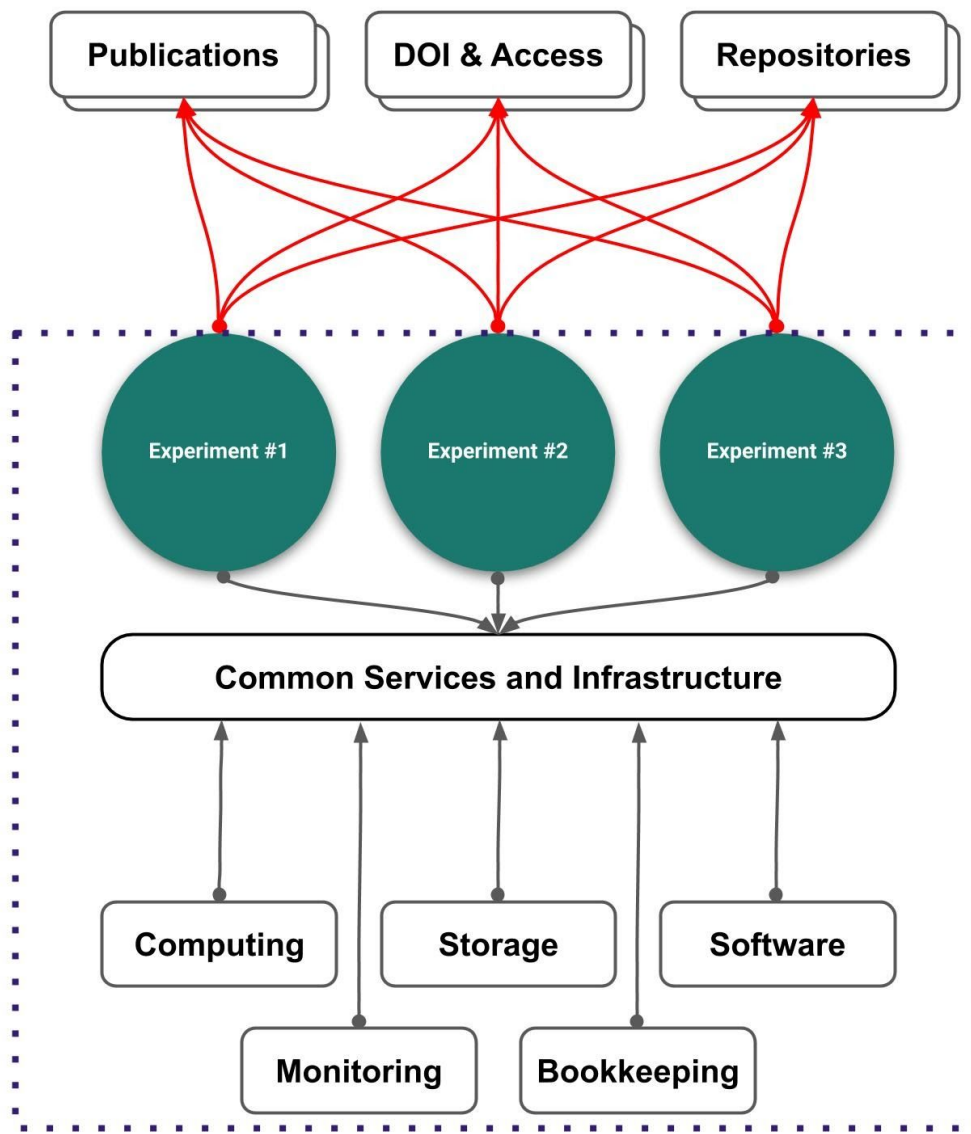

**Figure 7**: A simplified view of multiple experiments and the ways they can access to services and tools through a common Service Provider Platform in the areas of SC and IT highlighting the area of usage of Initial-continuous services.

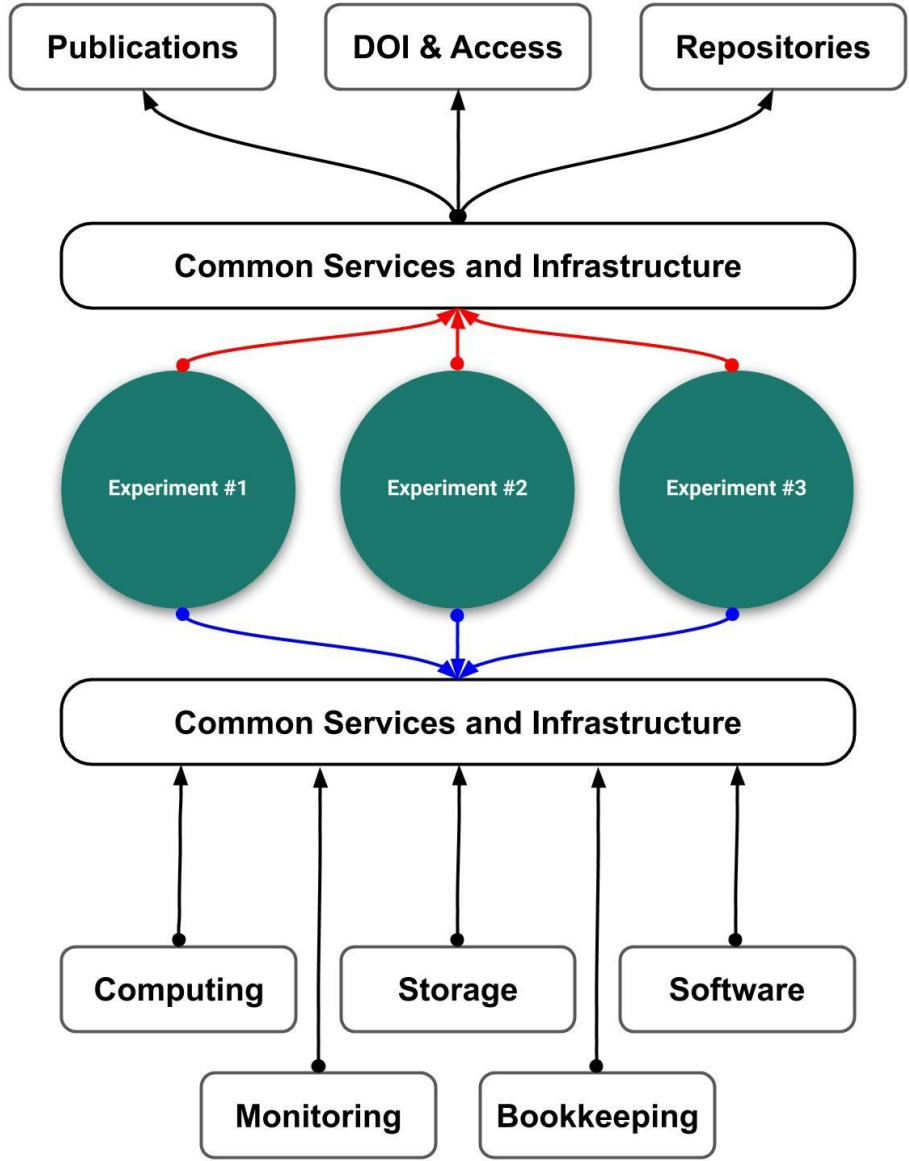

**Figure 8**: A simplified view of multiple experiments and the ways they can access to services and tools through common Service Provider Platforms in the areas of SC and IT highlighting the area of usage of ALL needed services.

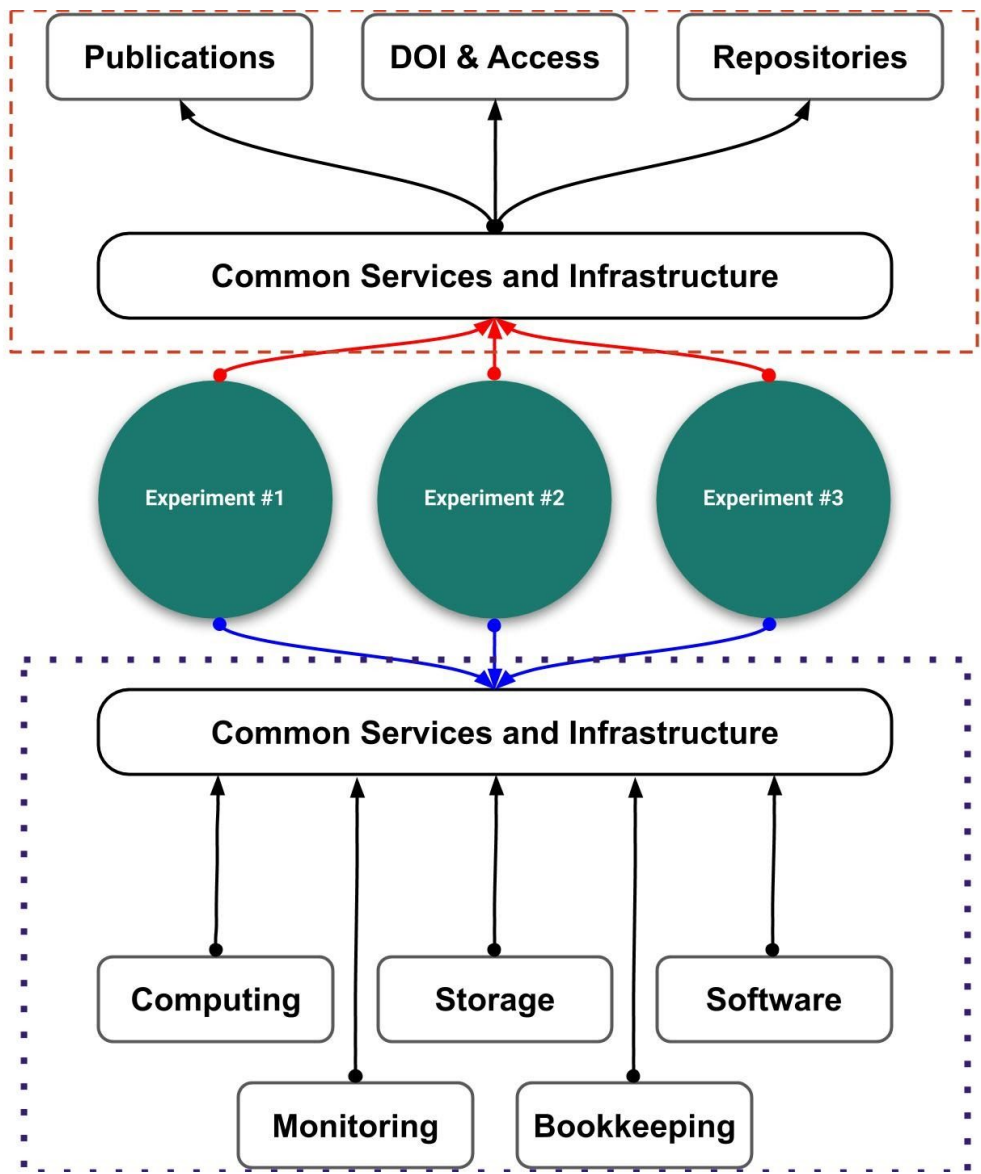

**Figure 9**: A simplified view of multiple experiments and the ways they can access to services and tools through common Service Provider Platforms in the areas of SC and IT highlighting the area of usage of ALL needed services. Adding the two main areas to focus on.

**In this vision, the reproducibility of the science by use of Open Data access is a relevant (if not the most important) goal also. Using services developed on public and/or generic platforms.** A current new field for capacity build and jobs/employment creation is easily imaginable.

As mentioned above, using global industrials standards by design. Together with education and discipline will be crucial at the moment of keeping those protocols, but it will pay off in the long run. Something senior scientists know well.

The obvious connection with the overall development of the region in the form of the creation of high-aggregated value products, leaving behind the economies of raw material.

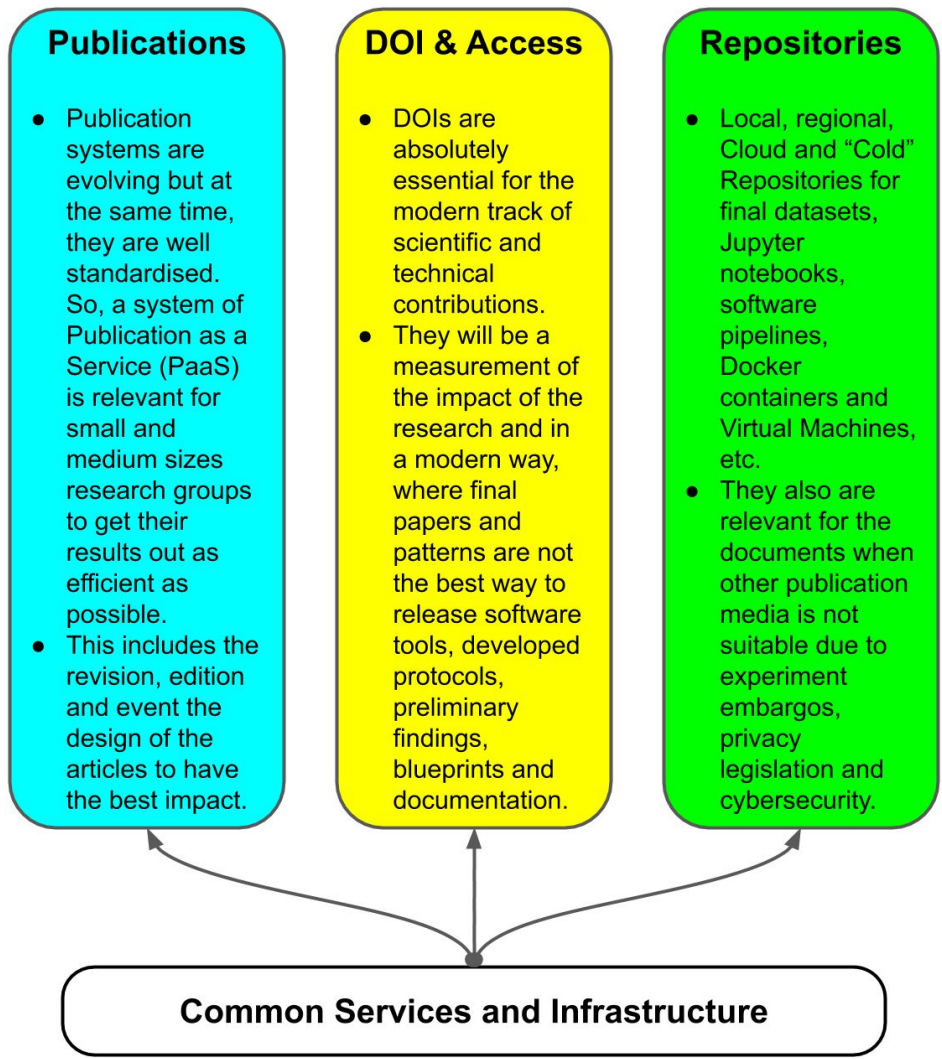

**Figure 10**: A non-exhausted set of components and resources that help to exemplify the vision of standard services for needs and activities that tend to be generic among different experiments and research activities in the sciences, particularly in physics.

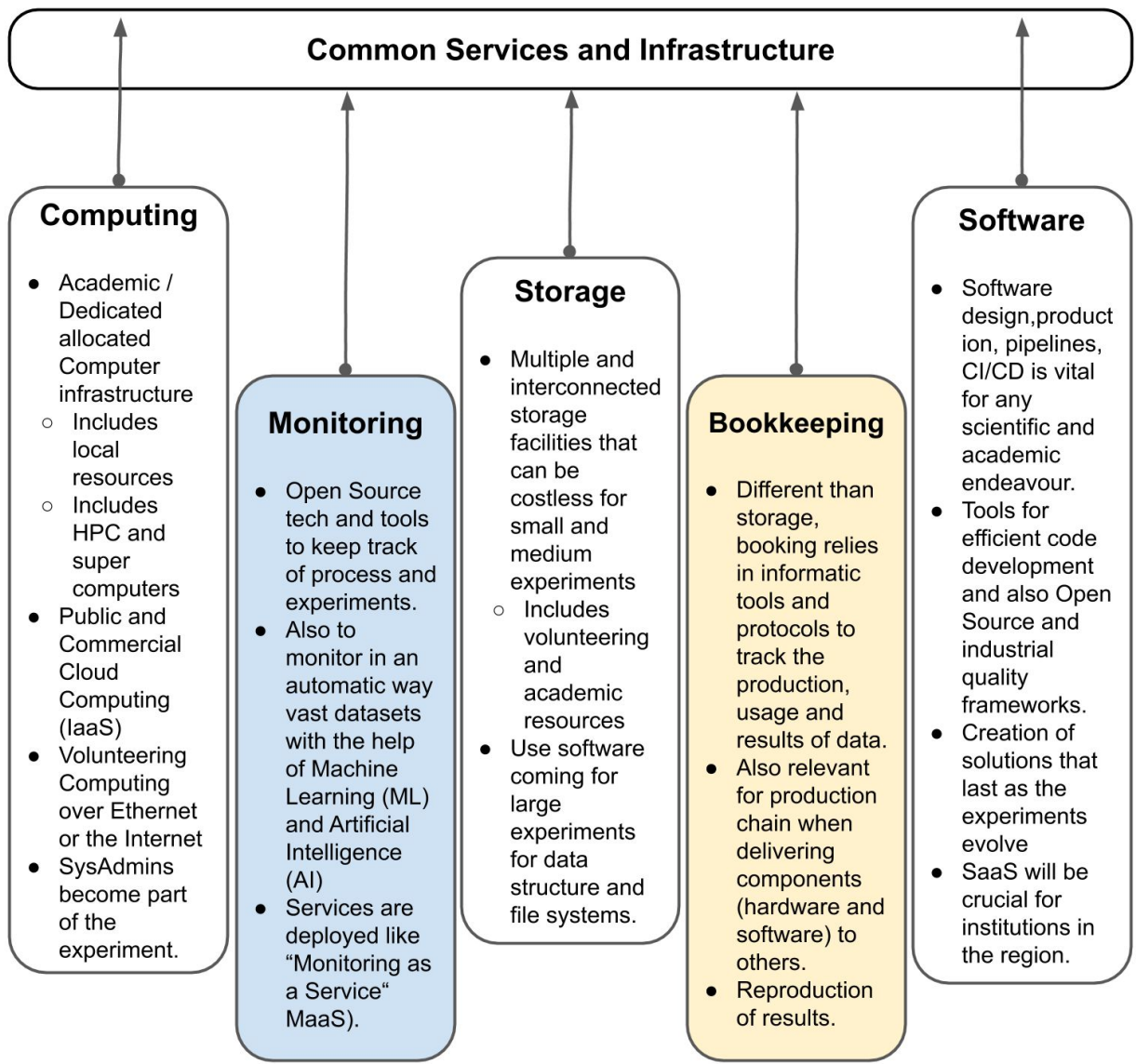

**Figure 11**: A non-exhausted set of components and resources that help to exemplify the vision of standard services for needs and activities that tend to be generic among different experiments and research activities in the sciences, particularly in physics.

The search for services providers should also be done in regional and local companies and services providers and boost the creation of the missing ones to boost the integration to multinational projects, and it matches the culture and requirements of funding agencies strategies like those of the European Union.

Also, encourage the creation and enhance resources to make -at least partially- self-sufficient several labs and institutions. For example, those related to services like food quality control, metrology, isotopes creation for medicine, among others. But in this proposal, those services can be provided in software and hardware development, quality control, consultancy, industrial assessments, etcetera.

# Preliminary remarks

It is important to mention that the deployment, support, consultancy and delivery of computational services (notably Cloud Computing, Big Data, storage, DOI & licence, ML and AI as a Service) are a very profitable and growing industry by itself.

Supported in the Open Source and Open Access culture, this kind of activities can be (and they will be) a vehicle for the development of the region in terms of the job creation, capacity-building and industrial production.

We should keep in mind that as a society, we must move forward in the scientific regime by the hand of the technological systems and the positive impact of both in the standard of living of the society. This includes the protection of natural environments and the fight against climate change.

We will reach those objectives with the creation of expertise and innovations in SC+IT industries and services for sciences, public and private sectors. Helping to shift our economies away from the traditional production of raw material to the creation of high-aggregated value products for their habitats and other markets worldwide.